\documentclass[pdftex,twocolumn,epjc3]{svjour3}      
\RequirePackage{amsmath}

\newcommand{\be}{\begin{equation}}
\newcommand{\ee}{\end{equation}}
\newcommand{\ba}{\begin{eqnarray}}
\newcommand{\ea}{\end{eqnarray}}
\newcommand{\nn}{\nonumber\\}

\RequirePackage{graphicx}
\RequirePackage{mathptmx}      
\RequirePackage{flushend}
\RequirePackage[numbers,sort&compress]{natbib}
\RequirePackage[colorlinks,citecolor=blue,urlcolor=blue,linkcolor=blue]{hyperref}

\begin{document}
	\title{Passage of heavy quarks through the fluctuating hot QCD medium}
	
	\author{Mohammad Yousuf Jamal\thanksref{e1,addr1}	\and
		Bedangadas Mohanty\thanksref{addr1} }
	
	\thankstext{e1}{mohammad.yousuf@niser.ac.in}
	\institute{School of Physical Sciences, National Institute of Science Education and Research, HBNI, Jatni-752050, India \label{addr1}}
	\maketitle
	
	\begin{abstract}
The change in the energy of the moving heavy (charm and bottom) quarks due to field fluctuations present in the hot QCD medium has been studied. A finite quark chemical potential has been considered while modeling the hot QCD medium counting the fact that the upcoming experimental facilities such as Anti-proton and Ion Research (FAIR) and Nuclotron-based Ion Collider fAcility (NICA) are expected to operate at finite baryon density and moderate temperature. The effective kinetic theory approach has been adopted where the collisions have been incorporated using the well defined collisional kernel, known as Bhatnagar-Gross-Krook (BGK). To incorporate the non-ideal equations of state (EoSs) effects/ medium interaction effects, an extended effective fugacity model has been adopted. The momentum dependence of the energy change due to fluctuation for the charm and bottom quark has been investigated at different values of collision frequency and chemical potential. The results are exciting as the heavy quarks are found to gain energy due to fluctuations while moving through the produced medium at finite chemical potential and collision frequency.
	\end{abstract}
	
\section{Introduction}
The hot QCD medium produces at various experimental facilities provides a basis that helps in understanding the different phases of the early Universe. Especially, it helps in studying the Universe of the age of a few microseconds where the Quark-Gluon Plasma (QGP) phase is expected to exist. This medium is created by colliding the heavy-ions at ultra-relativistic speed in the laboratory that mimics the BigBang (MiniBang). The major hindrance in its study is its small size and short-lived nature. Therefore, to study such a  medium, one depends on the observed signatures at the detector end. One of the most prominent signatures is the suppression/enhancement in the yield of high $p_T$ hadrons mainly caused due to the energy loss/gain of heavy quarks while passing through the hot QCD medium created in heavy-ion collision (HIC). Moreover, the upcoming experimental facilities such as FAIR, NICA, {\it etc}, are expected to explore the QCD phases at finite baryon density and moderate temperature. Hence, one needs to incorporate the finite baryon chemical potential to analyze this medium.

The study regarding the change in the energy of the heavy quarks started back in the 1980s~\cite{bjorken1982energy, Thoma:1990fm}. Several articles are present in the literature that contain invaluable information about the energy loss/gain of heavy quarks moving in the hot QCD medium through several processes and also by different approaches~\cite{Braaten:1991jj, Mrowczynski:1991da, Thomas:1991ea, Koike:1992xs, Romatschke:2004au, Baier:2008js, Carrington:2015xca, Majumder:2010qh, Mustafa:1997pm, Jeon:2003gi, Gyulassy:1999zd, Mustafa:2003vh, DuttMazumder:2004xk, Chakraborty:2006db, Adil:2006ei, Jamal:2019svc, Fadafan:2012qu}.In the current manuscript, the main aim is to study the change in the energy of moving heavy quarks (charm and bottom) due to fluctuations in the presence of small but finite quark chemical potential within the hot QCD medium.  Moreover, we intend to see how the presence of finite quark chemical potential in the medium, the collisional frequency of the medium constituents, and the non-ideal medium effects influence their energies. For that, we employ the effective kinetic theory approach,  considering the BGK- collisional kernel along with the extended EQPM ~\cite{Jamal:2020emj, chandra_quasi1, chandra_quasi2, Mitra:2017sjo}.

The manuscript is organized as follows. In section~\ref{sec:EL}, the formalism for the change in the energy of moving heavy quarks due to fluctuations at finite chemical potential is provided. Sections~\ref{el:RaD}, contains various observations and results of the energy change at different values of quark chemical potential, collision frequency, and the presence of the non-ideal medium effect. Section~\ref{el:SaF}, is dedicated to the summary and future possibilities of the present work.

\section{Energy change due to fluctuation }
\label{sec:EL}
The heavy quarks produced at the early stages after the HIC traverse through the produced medium almost as the independent degrees of freedom. Treating them as a classical particle, their motion within the medium can be described using Wong's equations~\cite{Wong:1970fu}. These equations are a set of classical equations of motion for a particle interacting with a chromo-dynamical field, $F_{a}^{\mu\nu}$, given in the Lorentz covariant form as,
\ba
\frac{dx^{\mu}(\tau)}{d\tau} &=& u^{\mu}(\tau),\nn
\frac{dp^{\mu}(\tau)}{d\tau} &=&g q^{a}(\tau)F^{\mu\nu}_{a}(x(\tau))u_{\nu}(\tau),\nn
\frac{dq^{a}(\tau)}{d\tau} &=& -gf^{abc}u_{\mu}(\tau)A^{\mu}_{b}(x(\tau))q_{c}(\tau),
\label{eq:wong}
\ea 
where, $q^a(\tau)$ is the quark's color charge and $g$, is the coupling constant. $\tau$,  $x^{\mu}\equiv X $, $u^{\mu}=\gamma(1,{\bf v})$ and $p^{\mu}(\tau)$ are the proper time, trajectory, four velocity  and four momentum of the heavy quark, respectively. Here, we have $\text N_{c}^{2}-1$ chromo-electric/magnetic fields. The structure constant, $f^{abc}$ belong to the ${\text {SU}({\text N_c})}$ gauge group and $A^{\mu}_{a}$ is the four potential. The expression of the energy change of heavy quarks can be obtained from Eq.~\ref{eq:wong} considering the  following two assumptions~\cite{Jiang:2016duz, Carrington:2015xca}. First, chosing the gauge condition $u_{\mu}A^{\mu}_{a} = 0$ which says that $q^{a}$ is independent of $\tau$. Second, the quark's momentum and energy evolve in time without changing much the magnitude of its velocity while interacting with the chromodynamic field. Solving Eq.~\ref{eq:wong} considering only the zeroth component, $\mu =0$ with $t=\gamma \tau$, we obtained,
\ba
\frac{\text{dE}}{\text{dt}}=g~q^a~\left< {\bf v}(t)\cdot {\bf E}^{a}(X(t))\right >,
\label{eq:el2}
\ea
where, $\left <.....\right >$ denotes the ensemble average.
The chromo-electric field in Eq. ~(\ref{eq:el2}) consists of the induced field as well as the spontaneously generated microscopic field due to the random function of position and time. Next, solving  $\mu \ne 0$ components in Eq.~(\ref{eq:wong}) we obtained,
\ba
\frac{{\text d{\bf p}}}{{\text {dt}}}=g~q^a \left[{\bf E}_{t}^{a}(X)+ {\bf v}\times {\bf B}_{t}^{a}(X)\right].
\label{eq:pt}
\ea
Using Eq. ~(\ref{eq:pt}) while considering $\left< E^a_iB^a_j\right >=0$ along with the fact that the mean value of the fluctuating part of the field equals zero, {\it i.e.,} $\left< {\tilde{E}}\right >=0$, Eq.~(\ref{eq:el2}) further reduces as ~\cite{Chakraborty:2006db}, 
\ba
\frac{{\text {dE}}}{{\text {dt}}}&=&\left< g~q^a {\bf v}\cdot {\bf E}^a\right > \nn &+&
g^2~\frac{q^a q^b}{E_0} \int_{0}^{t}dt_1\left< E^b_t (t_1)\cdot E^a_t(t)\right > \nn &+&
g^2~\frac{q^a q^b}{E_0} \int_{0}^{t}dt_1\int_{0}^{t}dt_2\bigg< \Sigma_j E^b_{t,j} (t_2)\cdot\frac{\partial}{\partial r_{j}} {\bf v}\cdot {\bf E}^a_t(t)\bigg >,
\label{el_full}
\ea
where, ${\bf v}=\frac{\bf p}{\sqrt{{\bf p}^2+M^2}}$ is the velocity of the heavy quark of mass $M$ and $E_0=\sqrt{{\bf p}^2+M^2}$ is the initial heavy quark energy. The Eq.~(\ref{el_full}) represents the full expression of the change in energy of heavy quarks where the first term corresponds to the change due to polarization. The other two terms correspond to the statistical change in the energy of the moving heavy quark in the medium due to the fluctuations of the chromo-electromagnetic fields as well as the velocity of the particle under the influence of this field. Precisely, the second term in Eq.~(\ref{el_full}) corresponds to the statistical part of the dynamic friction due to the space-time correlation in the fluctuations in the chromo-electrical field. Whereas, the third term corresponds to the average change in the energy of the moving heavy quark due to the correlation between the fluctuation in the velocity of the particle and the fluctuation in the chromo-electrical field in the plasma. The change in energy due to polarization has been studied earlier~\cite{Jamal:2020emj} and found that both charm and bottom quarks lose their energy while passing through the medium. Here, the main focus is to understand the effects on the change of energy of moving heavy quarks due to fluctuations in the medium. Therefore, to get the total contribution of energy exchange due to fluctuations, we shall only concentrate on the plus of the second and the third terms of Eq. \ref{el_full} that  can be further written as~\cite{Chakraborty:2006db},	

{\small	\ba
	\frac{\text{dE}}{\text{d {\bf x}}}&=&\frac{C_F \alpha _s}{\pi E_{0} \text{v}^4}
	\int^{k_{\infty} \text{v}}_{0} \omega^2d\omega \coth\left(\frac{\beta \omega }{2}\right)\frac{\text {Im}[\epsilon_L(\omega,k=\omega/\text{v})]}{|\epsilon_L(\omega,k=\omega/\text{v})|^2}\nn&+&
	\frac{2C_F \alpha _s}{\pi E_{0} \text{v}^2}
	\int^{k_{\infty}}_{k_0} dk k \int^{k}_{0}d\omega \coth\left(\frac{\beta \omega }{2}\right)\frac{\text {Im}[\epsilon_T(\omega,k)]}{|\epsilon_T(\omega,k)-k^2/\omega^2|^2},
	\label{eq:elfl}
	\ea} 
where, $C_F =4/3$ is the Casimir invariant in the fundamental representation of the $\text {SU}({\text N_c})$. The QCD running coupling constant, $\alpha_{s}(\mu_{q}, T)$ at finite chemical potential and temperature~\cite{ Mitra:2017sjo, Srivastava:2010xa} given as,

{\small	\ba
	\alpha_{s}(\mu_{q}, T)&=&\frac{g^2_{s}(\mu_{q}, T)}{4 \pi}\nn
	&=& \frac{6\pi}{\left(33-2N_f\right)\ln\left(\frac{T}{\Lambda_T} \sqrt{1+\frac{\mu^2_q}{\pi^2 T^2}}\right)}\nn
	&\times& \bigg(1-\frac{3\left(153-19N_f\right)}{\left(33-2N_f\right)^2}
	\frac{\ln \left(2\ln \left( \frac{T}{\Lambda_T} \sqrt{1+\frac{\mu^2_q}{\pi^2 T^2}}\right)\right)}{\ln \left( \frac{T}{\Lambda_T} \sqrt{1+\frac{\mu^2_q}{\pi^2 T^2}}\right)}\bigg).
	\ea}
The $\epsilon_{L}$ and $\epsilon_{T}$ are the longitudinal and transverse components of the medium dielectric permittivity discussed earlier in Ref.~\cite{Jamal:2019svc}. We shall briefly the major steps in the next subsection.
	\begin{figure*}
	\centering
	\includegraphics[height=5cm,width=7.60cm]{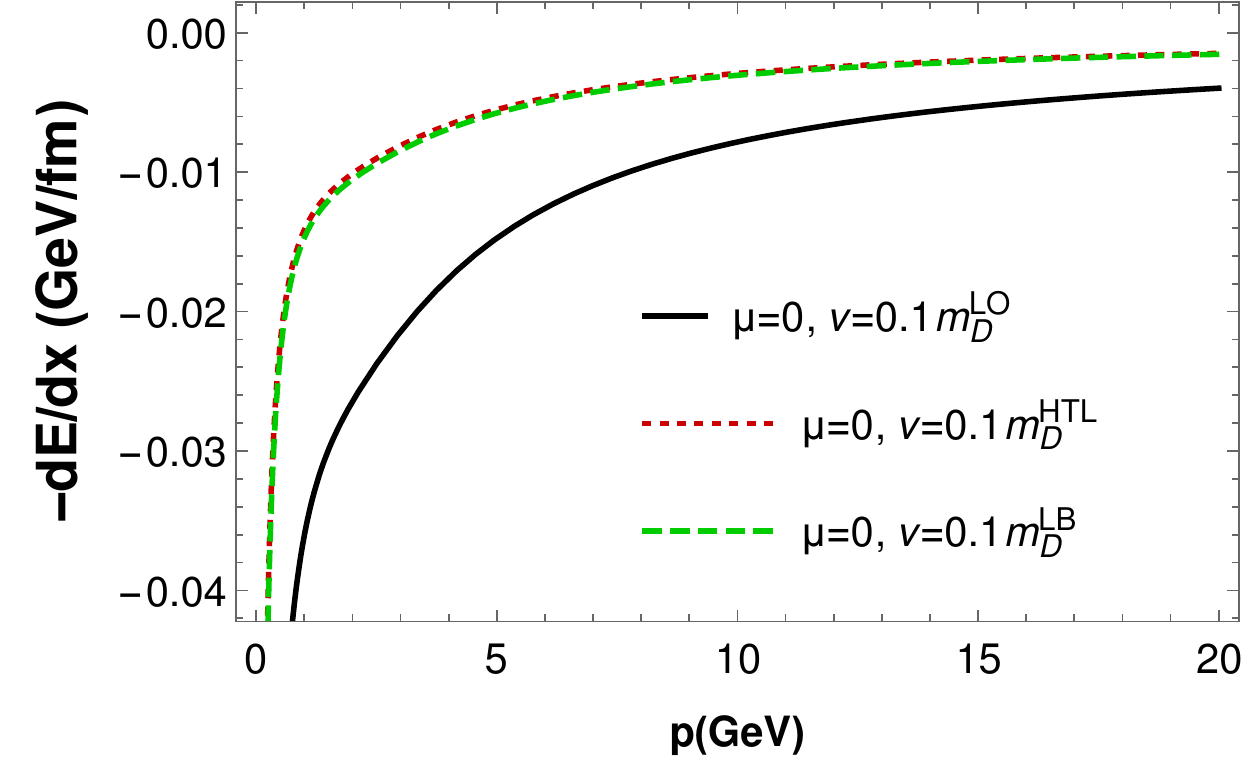}
	\hspace{3mm}
	\includegraphics[height=5cm,width=7.60cm]{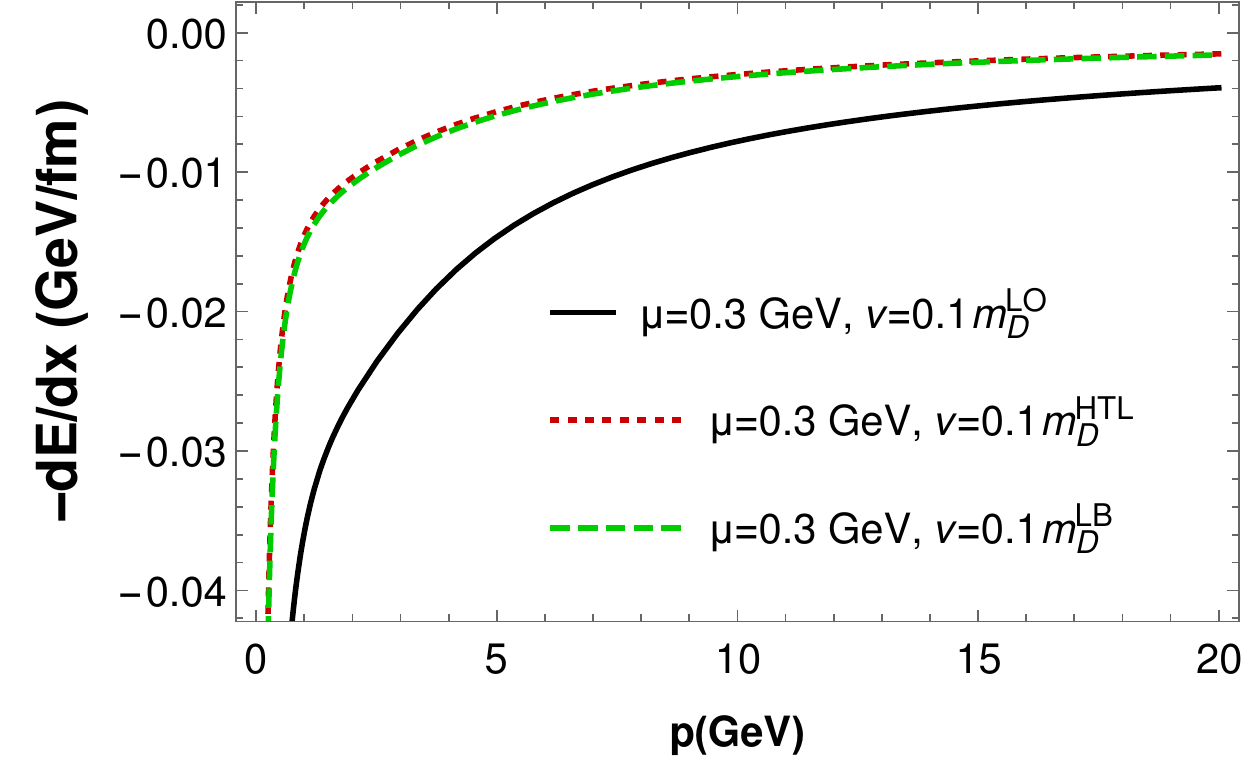}
	\caption{Energy change of charm quark due to fluctuation with various EoSs. In the left panel, $\mu =0$ and in the right panel, $\mu = 0.3~$GeV. }
	\label{fig:charm}
\end{figure*}	

\begin{figure*}
	\centering
	\includegraphics[height=5cm,width=7.60cm]{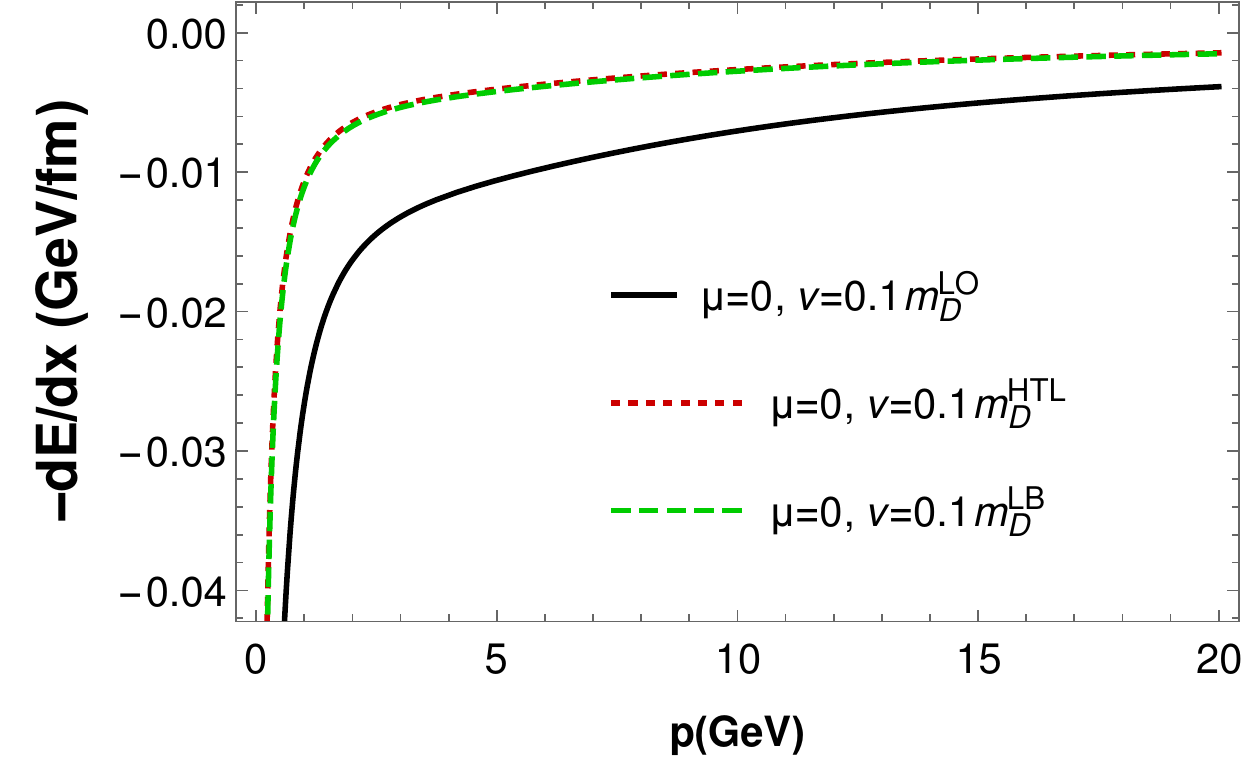}
	\hspace{3mm}
	\includegraphics[height=5cm,width=7.60cm]{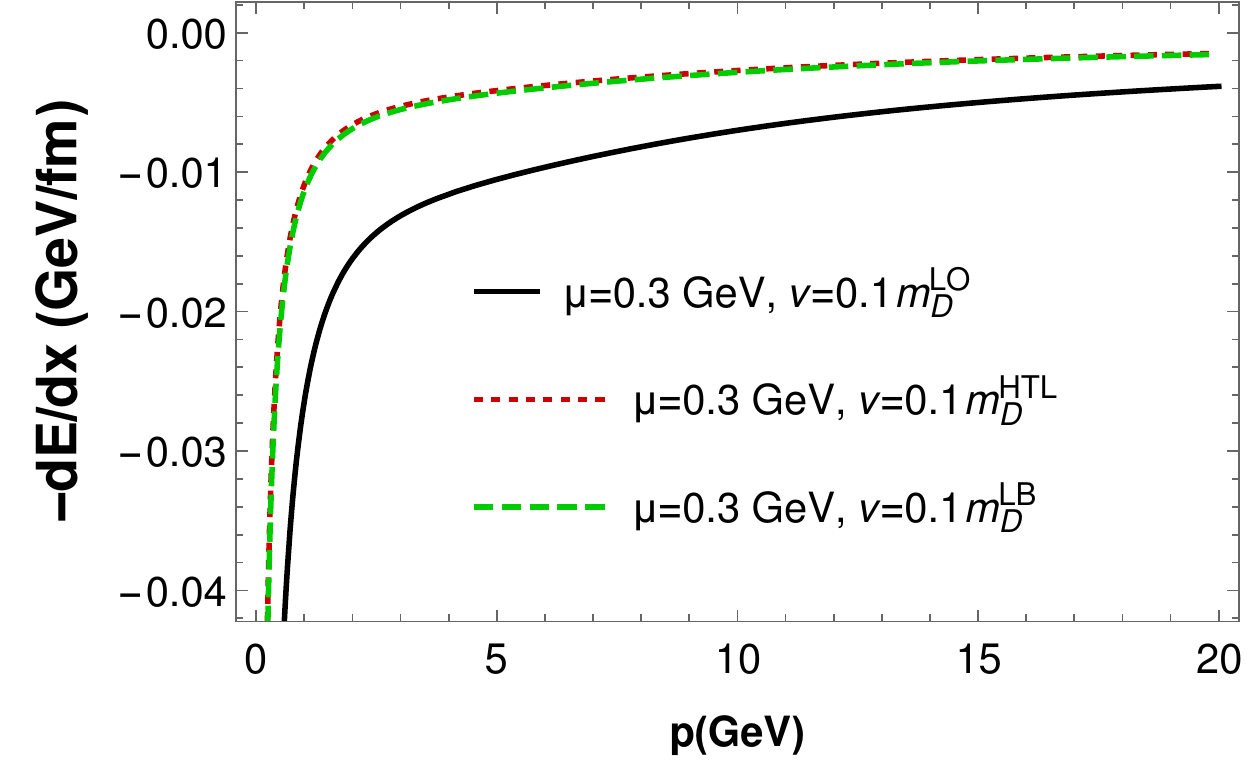}
	\caption{Energy change of bottom quark due to fluctuation with various EoSs. In the left panel, $\mu =0$ and in the right panel, $\mu = 0.3~$GeV. }
	\label{fig:bottom}
\end{figure*}
\subsection{Dieletric permittivity at finite quark chemical potential}
   For the isotropic hot QCD medium, the dielectric permittivity can be expanded in terms of its longitudinal and transverse projections as,
  \ba
  \epsilon^{\text{ij}}(K,\nu,T,\mu_q)= A^{ij}~\epsilon_T(K,\nu,T,\mu_q)+B^{ij}~\epsilon_L(K,\nu,T,\mu_q),
  \label{eq:eplt1}\nn
  \ea  
where, $K \equiv K^{\mu}=(\omega,{\bf k})$, $\mu_{q}$ represents the quark chemical potential and $\nu$, shows the collision frequency of the medium particles. The projection tensors are defined as, $A^{\text{ij}}=\delta ^{\text{ij}}-\frac{k^i k^j}{k^2}$, and $ B^{\text{ij}}=\frac{k^i k^j}{k^2}$. The permittivity tensor, $\epsilon^{ij}$ can be obtained from the polarisation tenser, $\Pi^{ij}$ as,
   \ba
   \epsilon^{ij}=\delta^{ij}-\frac{1}{\omega^2}\Pi ^{\text{ij}}.
   \label{eq:ep}
   \ea
where, $\Pi^{ij}$ can be derived from the induced current as,
   \ba
   \Pi^{ij}=\frac{\delta J^{i}_{a, ind}}{\delta A_{j, a}}.
   \label{eq:p}	 
   \ea
The current induced, $J_{ind, a}^{i}(X)$ inside the QCD medium due to the change in the particle distribution function is given as~ \cite{Mrowczynski:1993qm, Romatschke:2003ms, Jiang:2016dkf, Schenke:2006xu},
\ba
J_{ind,a}^{i}(X)&=&g\int\frac{d^{3}p}{(2\pi)^3} u^{i}\{2N_c \delta f^{g}_a(p,X)+N_{f}[\delta f^{q}_a(p,X) \nn
&-&\delta f^{\bar{q}}_a(p,X)]\}\label{indcurrent}.
\ea
Next, the change in the medium particles distribution functions, $\delta f^{i}$ can be obtained by solving the Boltzman-Vlasov transport equation that can be  written for each species as,
\ba
u^{\bar{\mu}}\partial_{\bar{\mu}} \delta f^{i}_a(p,X) + g \theta_{i}
u_{\bar{\mu}}F^{\bar{\mu}\bar{\nu}}_a(X)\partial_{\bar{\nu}}^{(p)}f^{i}(\mathbf{p})=C^{i}_a(\nu, p,X),  
\label{transportequation}
\ea
where, $\bar{\mu}$ and $\bar{\nu}$ are the Lorentz four indices (not to confuse with chemical potential and collision frequency). Index, $i$ represents the particle species ($i\in$ \{quarks, anti-quarks and gluons\}) and  $\theta_{i}\in\{\theta_g,\theta_q,\theta_{\bar{q}}\}$ have the values $\theta_{g} = \theta_{q} = 1 $ and $\theta_{\bar{q}}=-1$.  The partial four derivatives, $\partial_{\mu}$, $\partial_{\nu}^{(p)}$ correspond to the space and momentum, respectively. The collisional  kernel, ${C}^{i}_a(p,X)$ is considered here to be the BGK-type  ~\cite{Bhatnagar:1954} given as follows,
\ba
{C}^{i}_a(\nu, p,X)=-\nu\left[f^{i}_a(p,X)-\frac{N^{i}_a(X)}{N^{i}_{\text{eq}}}f^{i}_{\text{eq}}(|\mathbf{p}|)\right],\label{collision}
\ea
where, 
\ba
 f^{i}_a(p,X)=f^{i}(\mathbf{p})+\delta f^{i}_a(p,X),
\ea
are the distribution functions of quarks, anti-quarks and gluons, $f^{i}(\mathbf{p})$ is equilibrium part while, $\delta f^{i}_a(p,X)$ is the perturbed part of the medium particle distribution functions such that $\delta f^{i}_a(p,X)\ll f^{i}(\mathbf{p})$.   The collisions frequency,  $\nu$ is  considered here to be independent of momentum and particle species. The particle number, $N^{i}_a(X)$ and its equilibrium value, $N^{i}_{\text{eq}}$ are defined as follows,
\ba
N^{i}_a(X)=\int \frac{d^{3}p}{(2\pi)^3} f^{i}_a(p,X),
\ea
\ba
 N^{i}_{\text{eq}} = \int \frac{d^{3}p}{(2\pi)^3} f^{i}(\mathbf{p})\text{.}
\ea
 Solving Eq.~(\ref{indcurrent}) along with Eq.~(\ref{transportequation}) and Eq.~(\ref{eq:p}), in the Fourier space, we obtained,	
\ba
\Pi ^{ij}(K,\nu, \mu_q,T)&=& m_D^2(T,\mu_q)\int \frac{d\Omega }{4 \pi }u^{i} u^{l}\Big\{u^{j} k^{l}\nn
&+&\left(\omega -{\bf k}\cdot {\bf v}\right)\delta ^{lj}\Big\} D^{-1}\left(K,\nu \right),
\label{eq:pi}
\ea
where, $D\left(K,\nu \right)=\omega +i \nu-{\bf k}\cdot {\bf v}$. The squared Debye mass given as,
\ba
m^{2}_D(T,\mu_q) &=& 4\pi \alpha_s(T,\mu_q) \bigg(-2N_c \int \frac{d^3 p}{(2\pi)^3} \partial_p f_g(p)\nn &-& N_f \int \frac{d^3 p}{(2\pi)^3} \partial_p \left(f_q(p)+f_{\bar q}(p)\right)\bigg).
\label{eq:md} 
\ea    
From Eq.~(\ref{eq:eplt1}), ~(\ref{eq:ep}) and (\ref{eq:pi}) we have,
\ba
\epsilon _L(K,\nu,T,\mu_q)&=&1+m_D^2(T,\mu_q) \frac{2 k-(\omega +i \nu ) \ln \left(\frac{k+i \nu +\omega }{-k+i \nu +\omega }\right)}{k^2 \left(2 k-i \nu  \ln \left(\frac{k+i \nu +\omega }{-k+i \nu +\omega }\right)\right)}~\nn
\epsilon _T(K,\nu,T,\mu_q)&=&1-\frac{m_D^2(T,\mu_q)}{2~ \omega~k} \bigg[\frac{\omega +i \nu}{k}+\Big(\frac{1}{2}-\frac{(\omega +i \nu )^2}{2k^2}\Big)\nn
&\times& \ln \left(\frac{k+i \nu +\omega }{-k+i \nu +\omega }\right)\bigg].
\ea
In the next subsection, we shall discuss the inclusion of non-ideal hot QCD medium effects using lattice and HTL equations of states (EoSs) along with the finite quark chemical potential. 
	\begin{figure}
	\centering
	\includegraphics[height=5cm,width=8.50cm]{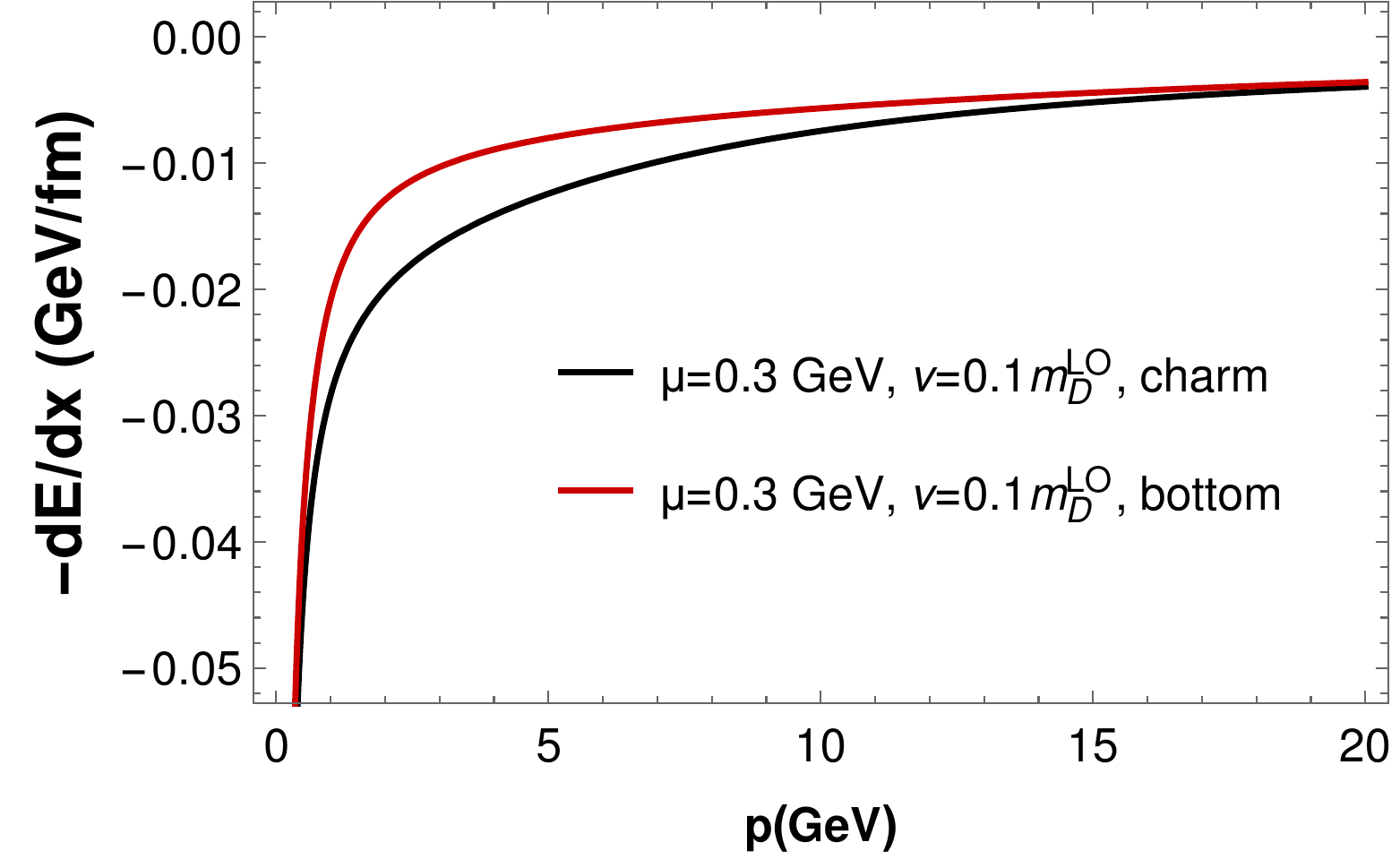}
	\caption{Energy change of charm and bottom quarks at fixed $\mu =0.3~$GeV and $\nu =0.1m_D^{LO}~$ considering only the leading order case.}
	\label{fig:CB}
	\end{figure}
	\begin{figure*}
	\centering
	\includegraphics[height=5cm,width=8.50cm]{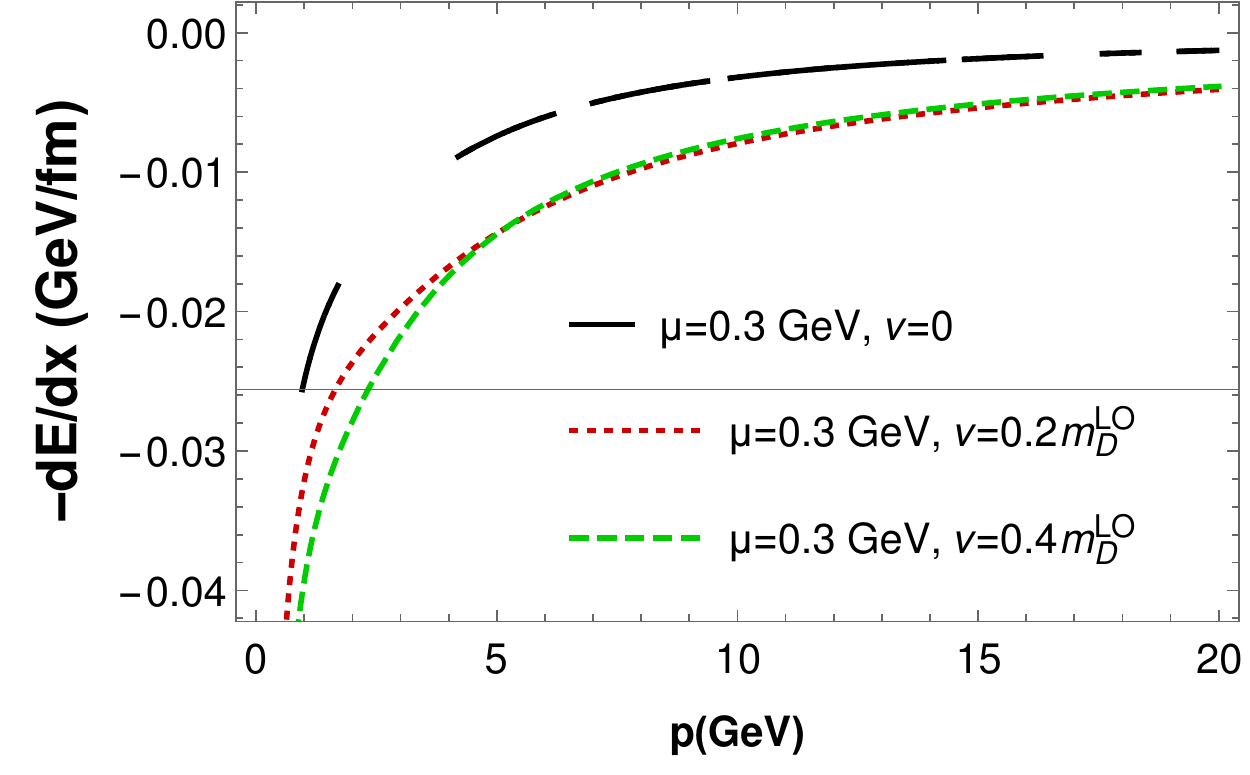}
	\includegraphics[height=5cm,width=8.50cm]{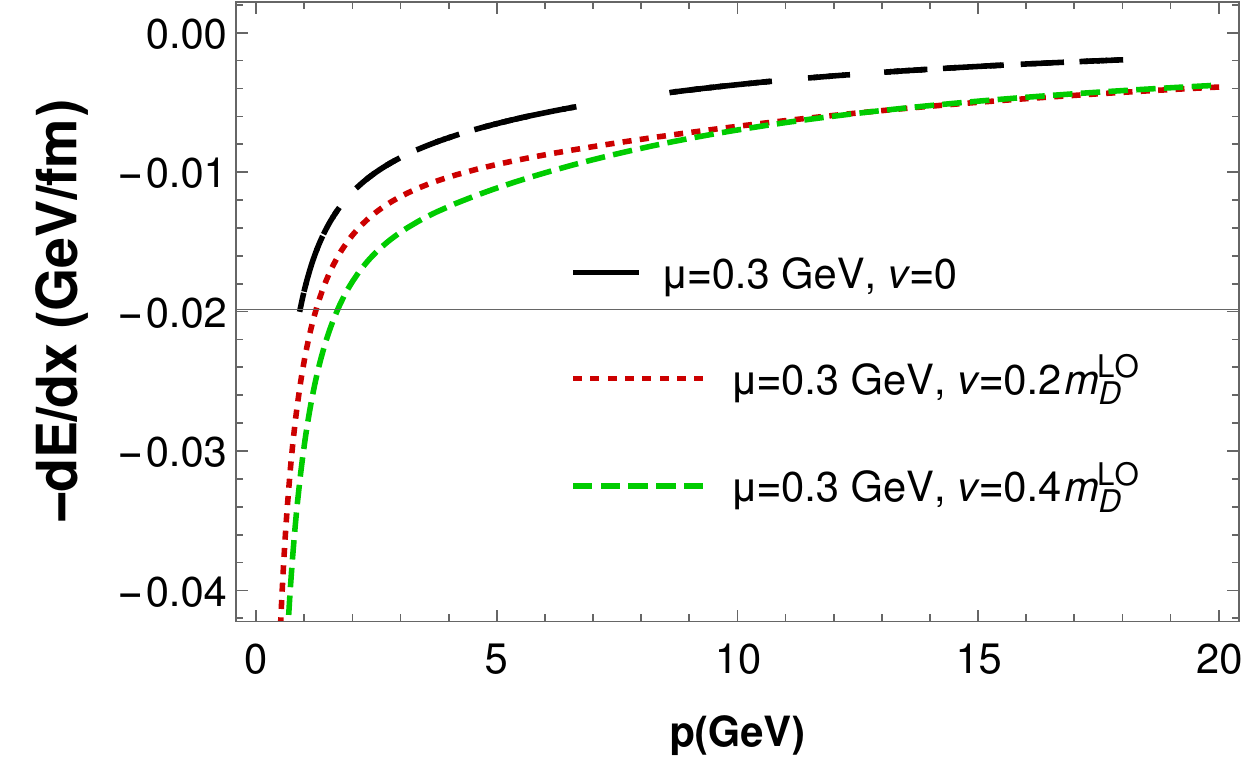}
	\caption{Energy change of charm quark (left panel) and bottom quark (right panel) at fixed $\mu =0.3~$GeV but different collisional frequencies considering only the leading order case.}
	\label{fig:cb}
\end{figure*}

\subsection{Extended EQPM to incorporate the medium interaction and the finite quark chemical potential }
The extended EQPM has been employed to incorporate the finite quark chemical potential along with the medium interaction effects (extended with the consideration of finite quark chemical potential, for details see Ref.~\cite{Mitra:2017sjo}). The model maps the hot QCD medium interaction effects present in the hot QCD EoSs either computed within perturbative QCD (pQCD) or lattice QCD simulations into the effective equilibrium distribution functions for the quasi-partons, $ f_{eq}^{i}\equiv \lbrace f_{g}, f_{q},f_{\bar q} \rbrace$, that in turn, describes the strong interaction effects in terms of effective fugacities, $z_{g,q}$~\cite{chandra_quasi1, chandra_quasi2}. The hot QCD EoSs described here are the recent ($2+1$)-lattice EoS from hot QCD collaboration ~\cite{bazabov2014}, and the 3-loop HTL perturbative EoS that has been computed by N. Haque {\it et,  al.} ~\cite{nhaque, Andersen:2015eoa} which agrees reasonably well with the lattice results~\cite{bazabov2014, fodor2014}. The EQPM has already been successfully applied to study the various aspects of hot QCD medium ~\cite{ Jamal:2017dqs, Kurian:2019nna, Jamal:2018mog, Agotiya:2016bqr, Kumar:2017bja, YousufJamal:2018ucf, Jamal:2020hpy}. 
In the present case, at finite quark chemical potential, the momentum distributions of gluon, quark, and anti-quark are given as,
\ba
\label{eq1}
f_{g}&=& \frac{z_{g}\exp[-\beta E_g]}{\bigg(1- z_{g}\exp[-\beta E_g]\bigg)},\nn
f_{q}&=& \frac{z_{q}\exp[-\beta ( E_q-\mu_q)]}{\bigg(1+ z_{q}\exp[-\beta ( E_q-\mu_q)]\bigg)},\nn 
f_{\bar q}&=& \frac{z_{q}\exp[-\beta ( E_q+\mu_q)]}{\bigg(1+ z_{q}\exp[-\beta ( E_q+\mu_q)]\bigg)},
\ea
where, $E_{g}=|{\bf p}_{g}|$ for the gluons and, $\sqrt{|{\bf p}_q|^2+m_q^2}$ for the quark ($m_q$, denotes the mass of the light quarks). The fugacity parameter, $z_{g/q}\rightarrow 1$ as temperature $T\rightarrow \infty$.
Since the model is valid only in the deconfined phase of QCD ({\it i.e., }beyond $T_c$, $T_c$ being the critical temperature), the masses of the light quarks can be neglected. Next, solving Eq.~(\ref{eq:md}), considering the distribution functions given in Eq.~(\ref{eq1}) within the limit, $\frac{\mu^2_q}{T^2}\ll1$ we obtained,
\ba
m^{2, EoS}_D(T,\mu_q) &=& 4\pi \alpha_s(T,\mu_q)T^2 \bigg[\Big(\frac{2N_c}{\pi^2}PolyLog[2,z_g^{EoS}]\nn
&-&\frac{2N_f}{\pi^2}PolyLog[2,-z_q^{EoS}]\Big)+\frac{\mu_q^2}{T^2}\frac{N_f}{\pi^2}\frac{z_q^{EoS}}{1+z_q^{EoS}}
\bigg],\nn
\label{eq:mde}
\ea
In the high temperature limit, $z_{g,q}\rightarrow 1$ and Eq.~(\ref{eq:mde}) reduces to leading order HTL expression,
\ba
	m^{2}_D(T,\mu_q) &=& 4\pi \alpha_s(T,\mu_q)T^2 \bigg[\frac{N_c}{3}+\frac{N_f}{6}
	+\frac{\mu_q^2}{T^2}\frac{N_f}{2\pi^2}
	\bigg].
\label{eq:mdh}
\ea
From Eq.~(\ref{eq:mde}) and Eq.~(\ref{eq:mdh}), one can obtain the effective coupling due to medium interaction effects through various EoSs at finite chemical potential as,
\ba
	\alpha_s^{EoS}(T,\mu_q) &=&\frac{\alpha_s(T,\mu_q)}{\Big(\frac{N_c}{3}+\frac{N_f}{6}\Big)
		+\frac{\mu_q^2}{T^2}\frac{N_f}{2\pi^2}}  \bigg[\Big(\frac{2N_c}{\pi^2}\text {PolyLog}[2,z_g^{EoS}]\nn
	&-&\frac{2N_f}{\pi^2}\text {PolyLog}[2,-z_q^{EoS}]\Big)
	+\frac{\mu_q^2}{T^2}\frac{N_f}{\pi^2}\frac{z_q^{EoS}}{1+z_q^{EoS}}
	\bigg].\nn
	\label{eq:ec}
\ea
Let us summarize here the inclusion of various observables in this investigation:\\
(i) The frequency, $\nu$ of the medium particle collision enters using the BGK-collisional kernel that in turn appeared in the dielectric permittivity.\\
(ii) The finite chemical potential is introduced in the analysis in two different ways, that is, the strong coupling and the medium quark/anti-quark distribution functions. \\
(iii) The non-ideal medium interaction effects incorporated using the fugacity parameter in the gluon, quark and anti-quark distribution functions.   

Next, we shall discuss the various results regarding the change in energy of the moving charm and bottom quarks due to fluctuations inside the medium considering the above mentioned observables.

\section{Results and discussion}
\label{el:RaD}
The change in the energy of moving charm and bottom due to fluctuation has been plotted against their momenta considering the presence of finite quark chemical potential, $\mu_q$,  collision frequency, $\nu$, and non-ideal medium effects (EoSs). To do so, Eq.(\ref{eq:elfl}) has been solved numerically. To perform the numerical integration, the lower limit is taken as $k^{EoSs}_{0}=0$. The upper limits, $k_{\infty}$ has an ultraviolet cutoff of the order of the Debye mass {\it i.e.,} $k^{EoSs}_{\infty}\sim m^{EoSs}_{D}(T)$ for each EoS. It is to note that the results obtained here have a contribution to the change in the heavy-quark energy arising from soft-momentum exchange processes. The hard collisions and medium-induced gluon radiation are not considered here as these are beyond the scope of the current analysis. Here, we worked at temperature, $T=2~T_c$, $T_c=0.17 $~GeV, $N_c=3$, $N_f=3$. The different values of collision frequency have been taken in the order of screening mass that could be taken maximum up to $\nu=0.62m_D$ as shown in Ref.~\cite{Schenke:2006xu}. The quark chemical potential is taken within the limit $\frac{\mu_q}{T}<1$. In all the figures, the y-axis of the plots show the energy loss heavy quark per unit length ({\it i.e.,} $-\frac{{\text {dE}}}{{\text {dx}}}$ in GeV/fm) and the x-axis show their corresponding momenta (GeV). The results for the ideal case or the leading order denoted as `LO' and the non-ideal cases, {\it viz.,} $(2+1)-$ lattice EoS and 3-loop HTL EoS denoted as `LB' and  `$\mathrm{HTL_{PT}}$', respectively in the plots.  

In Fig.~\ref{fig:charm}, the plots are shown for the energy change of charm quark at $\nu=0.1 m_D^{EoSs}$ and $\mu_q=0$ (left panel) and $\mu_q=0.3$~GeV (right panel). The same for the bottom quark is shown in Fig.~\ref{fig:bottom}. It has been observed that in the case of fluctuation, both the heavy quarks gain energy as the plotted energy loss is appeared to be negative. As the momentum going towards higher values, the gain moves towards the smaller values and then reaches saturation. Although, the results from $(2+1)$- Lattice and 3-loop HTLpt are almost overlapping. But as compared to the LO case, the result using non-ideal EoSs are found to be suppressed up to $50\%$. Furthermore, in the presence of chemical potential, $\mu=0.3~$GeV shows less gain than vanishing chemical potential, $\mu=0$. Although, the presence of finite but small quark chemical potential does not much affect the results as it reduces the gain maximum up to $5\%$.   

 The EoSs effects are quite clear in Fig.~ \ref{fig:charm} and ~\ref{fig:bottom}, thus to avoid the bulk we will now only focusing on LO case. Fig.~\ref{fig:CB} represents a comparison between change of energy of charm and bottom quark at $\mu=0.3~$GeV and $\nu=0.1m_D^{LO}$. It has been observed that the charm quark gain more energy as compared to the bottom quark although they both reaches to the same saturation values at high momentum. This is because, at very high momentum, ${\text v} \rightarrow 1$, and hence, their masses do not affect the results.

In Fig.~\ref{fig:cb}, only the LO results have been shown for charm (left panel) and bottom (right panel) at fixed $\mu$ but different $\nu$, $\nu =0,~ 0.2~m_D^{\text LO},~ 0.3~m_D^{\text LO}$ to make a visible effects of collision frequencies. It has been noticed that both the quarks gain less energy when the medium is considered to be collisionless{\it i.e.,} $\nu =0$ as compared to the collisional cases, $\nu \neq 0$. Although the effect of collision at $\nu=0.2m_D^{\text LO}$ and $\nu=0.4m_D^{\text LO}$ do not differ much. But at low momentum both the quarks are gaining more energy at $\nu=0.4m_D^{\text LO}$ than at $\nu=0.2m_D^{\text LO}$.

Next, we shall summarise the analysis and discuss the future extension of the current project. 	
	
\section{Summary and future aspects}
	\label{el:SaF}
	
The formalism for the change in the energy of the moving heavy quarks due to fluctuations inside the hot QCD medium has been presented. The impact of small but finite quark chemical potential, collision frequency, and non-ideal medium effects has also been studied. Effective semi-classical transport theory has been adopted throughout. The collision effects have been incorporated using the BGK-collisional kernel. For the inclusion of non-ideal interaction effects, as well as the finite quark chemical potential, extended EQPM has been employed. It has been observed that both the heavy quarks gain energy due to fluctuations while traversing through the hot QGP medium even in the presence of chemical potential and collisions in the medium. Although, the gain saturates at high momentum. The non-ideal medium interaction effects are found to suppress the gain whereas with the increase in collision frequency the gain increases. For the same values to chemical potential and collision frequency, the charm quark is found to gain more energy than the bottom quark. Moreover, the small quark chemical potential caused some reduction in the gain but does not affect the results much. The higher values of chemical potential may affect the observed results appreciably but is beyond the scope of the current analysis.  Therefore, in the near future, the aim is to develop a formalism for $\frac{\mu_q}{T}>1$.	 
	
Apart from this, the future extension of the current work would be the inclusion of magnetic field effects, momentum anisotropy as well as viscous effects while modeling the hot QCD/QGP medium. A full analysis considering both the polarization as well as fluctuation is also important within the consideration of mentioned above observables. Furthermore, the observed results may have a significant impact on the experimental observables such as the nuclear modification factor $R_{AA}$ of heavy mesons for the nucleus-nucleus and/or p-nucleus collisions. Therefore, an immediate future extension includes the investigation of $R_{AA}$ in the present context.

\section{Acknowledgements}
M. Y. Jamal acknowledges NISER Bhubaneswar for providing a postdoctoral position. We would like to acknowledge the people of INDIA for their generous support for the research in fundamental sciences in the country.

\end{document}